\documentclass[sigconf, screen=true, review=false, anonymous=false, authordraft=False]{acmart}

\usepackage[yyyymmdd]{datetime}

\usepackage{tabularx}
\usepackage{dcolumn} 
\newcolumntype{d}[1]{D{.}{.}{#1}}

\usepackage{subcaption}

\usepackage{todonotes}
\let\xtodo\todo
\renewcommand{\todo}[1]{\xtodo[inline,color=green!50]{#1}}

\definecolor{burgundy}{RGB}{144,0,32}

\AtBeginDocument{%
  }

\copyrightyear{2024} 
\acmYear{2024} 
\setcopyright{rightsretained} 
\acmConference[CHI EA '24]{Extended Abstracts of the CHI Conference on Human Factors in Computing Systems}{May 11--16, 2024}{Honolulu, HI, USA}
\acmBooktitle{Extended Abstracts of the CHI Conference on Human Factors in Computing Systems (CHI EA '24), May 11--16, 2024, Honolulu, HI, USA}
\acmDOI{10.1145/3613905.3636286}
\acmISBN{979-8-4007-0331-7/24/05}

\begin{document}


\title[PhysioCHI]{PhysioCHI: Towards Best Practices for Integrating Physiological Signals in HCI}


\settopmatter{authorsperrow=4}

\author{Francesco Chiossi}
 \orcid{0000-0003-2987-7634}
\affiliation{%
  \institution{LMU Munich}
  \postcode{80337}
  \country{Germany}
}
\email{francesco.chiossi@lmu.de}

\author{Ekaterina R. Stepanova}
\orcid{0000-0002-1983-1512}
\affiliation{%
  \institution{Simon Fraser University}
  \country{Canada}}
\email{katerina_stepanova@sfu.ca}

\author{Benjamin Tag}
\orcid{0000-0002-7831-2632}
\affiliation{%
  \institution{Monash University}
  \city{Melbourne}
  \country{Australia}}
\email{benjamin.tag@monash.edu}

\author{Monica Perusqu\'ia-Hern\'andez}
\orcid{0000-0002-0486-1743}
\affiliation{%
  \institution{Nara Institute of Science and Technology (NAIST)}
  \city{Nara}
  \country{Japan}}
\email{perusquia@ieee.org}

\author{Alexandra Kitson}
\orcid{0000-0003-3479-5297}
\affiliation{%
  \institution{Simon Fraser University}
  \city{Surrey}
  \country{Canada}}
\email{akitson@sfu.ca}

\author{Arindam Dey}
\orcid{0000-0002-6638-2422}
\affiliation{%
  \institution{The University of Queensland}
  \city{Brisbane}
  \country{Australia }}
\email{a.dey@uq.edu.au}

\author{Sven Mayer}
\orcid{0000-0001-5462-8782}
\affiliation{%
  \institution{LMU Munich}
  \city{Munich}
  \postcode{80337}
  \country{Germany}}
\email{info@sven-mayer.com}

\author{Abdallah El Ali}
\orcid{0000-0002-9954-4088}
\affiliation{%
  \institution{Centrum Wiskunde \& Informatica}
  \city{Amsterdam}
  \country{Netherlands}}
\email{abdallah.el.ali@cwi.nl}

\renewcommand{\shortauthors}{Chiossi et al.}

\begin{teaserfigure}
  \includegraphics[width=\textwidth]{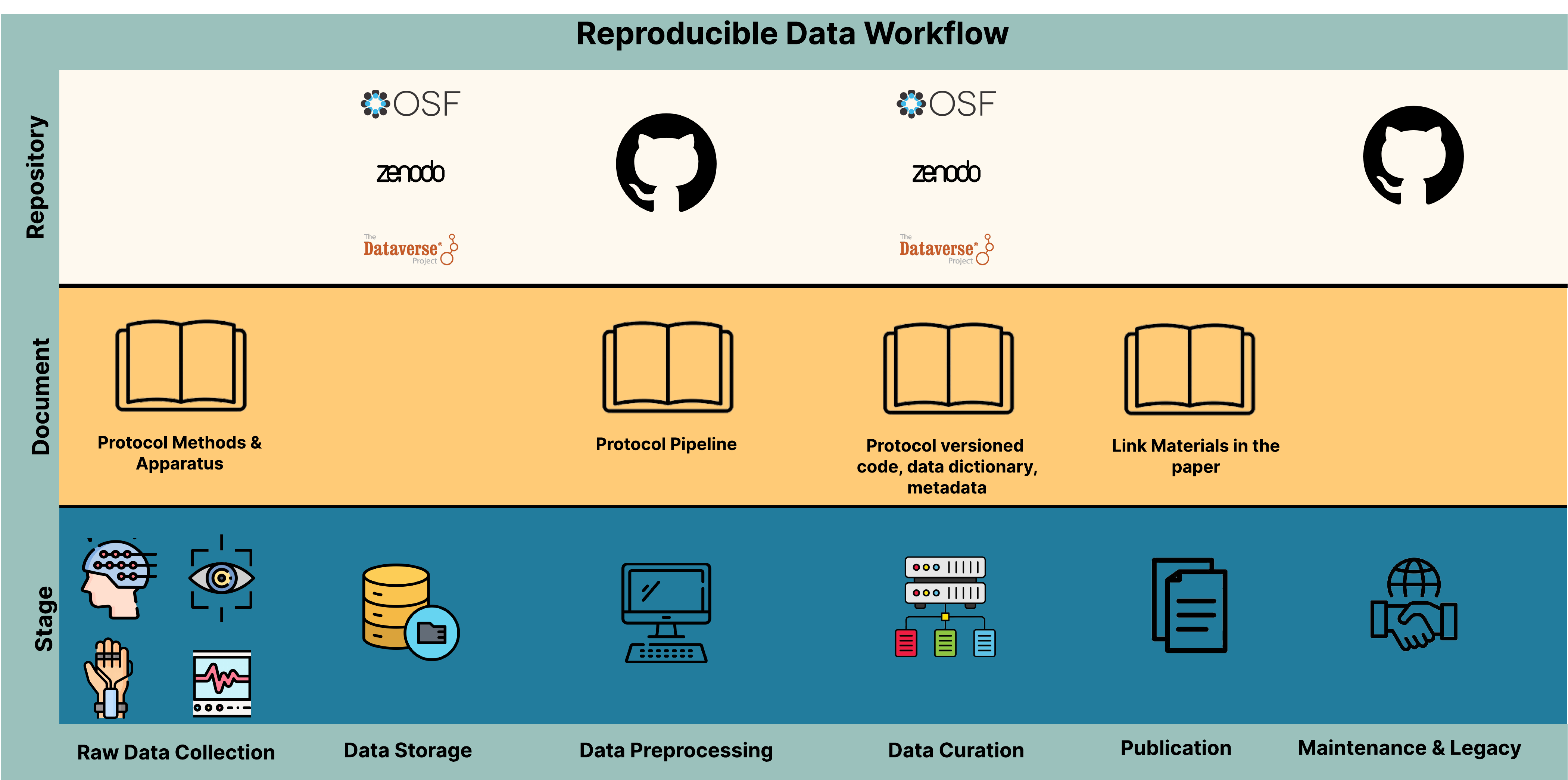}
\caption{\textit{Reproducibility Workflow for Physiological Signals in HCI.} Icons at the bottom represent the key steps in the workflow of a typical HCI user study, from inception to reporting. Book icons indicate components for reproducible best practices that map onto this workflow. Repository icons on top provide potential open-source platforms to share documented protocols.}
  \label{fig:teaser}
\end{teaserfigure}

\begin{abstract}
Recently, we saw a trend toward using physiological signals in interactive systems. These signals, offering deep insights into users' internal states and health, herald a new era for HCI. However, as this is an interdisciplinary approach, many challenges arise for HCI researchers, such as merging diverse disciplines, from understanding physiological functions to design expertise. Also, isolated research endeavors limit the scope and reach of findings. This workshop aims to bridge these gaps, fostering cross-disciplinary discussions on usability, open science, and ethics tied to physiological data in HCI. In this workshop, we will discuss best practices for embedding physiological signals in interactive systems. Through collective efforts, we seek to craft a guiding document for best practices in physiological HCI research, ensuring that it remains grounded in shared principles and methodologies as the field advances.
\end{abstract}




\maketitle

\section{Motivation}
In the human-computer interaction (HCI) community, we see a growing interest in integrating physiological signals (e.g., heart rate, breathing, electrodermal activity, brain activity, muscle tension) for usability evaluation and as input for interactive systems. \textbf{Physiological sensors} are devices that measure our physiological signals and, when analyzed, can provide insights into our physical and affective states~\cite{shu2018review}. Additionally, physiological sensors can be employed in physiologically-adaptive systems for tailoring interactions to users' states or prompt changes in the system to encourage a desirable user state~\cite{chiossi2022adapting, chiossi2023designing, chiossi2023adapting}. Here, the field is developing traction, and we foresee more growth as sensor technology becomes more viable and embedded in interactive systems~\cite{bernal2022galea, chiossi2023senscon}. A recent increase in review papers mapping out sections of the design domain with physiological sensors reveals the desire of the community to structure and better understand the dimensions of this emerging design space. For instance, \citet{moge2022shared} presented a systematic review on interpersonal biofeedback outlining the use of physiological signals in social interaction. Previously, \citet{prpa2020inhaling} presented an analysis of theoretical frameworks underlying the design of breath-responsive systems. In addition, \citet{yu2018biofeedback} reviewed biofeedback systems for stress management, identifying several challenges in the interpretation of signals, scalability of systems, and sparse evaluation. While all of these systematic revisions outlined biosignals' existing and future opportunities, others have identified methodological issues and a lack of shared best practices~\cite{babaei2021critique, putze2022understanding, treacy2015designing}. Specifically, integrating physiological signals in HCI necessitates a diverse skill set, from understanding physiological functioning and its associated psychological and cognitive processes to expertise in signal acquisition and processing, machine learning, and design, each a research area of its own.

The confluence of multiple disciplines presents a challenging learning curve for researchers and designers looking to incorporate physiological signals into studies and interactive systems. In contrast to building upon peers' work, the current landscape of physiological signals in HCI sees most research groups independently developing their datasets, analysis pipelines, and methodologies. Thus, HCI contributions to physiological computing are challenging as they must adhere to high standards from multiple disciplines (neuroscience, cognitive science, design). To meet such rigorous standards, HCI researchers must develop expertise in each discipline. There is a need to provide a shared framework and guidelines to support HCI research using physiological signals.

Stemming from the work in other disciplines~\cite{clayson2022open, poldrack2017scanning, govaart2022eeg}, HCI researchers have started to provide guidelines to support incorporating physiological signals. Here, \citet{babaei2021critique} evaluated practices for the recording and preprocessing of electrodermal activity (EDA). Similarly, \citet{putze2022understanding} have provided reporting practices for electroencephalographic signals (EEG) to support reproducible and reusable results. Finally, \citet{treacy2015designing} have evaluated experimental practices for fNIRS research in HCI and highlighted principles and patterns for effective brain-based adaptive interaction techniques. While there have been strides towards the establishment of reporting guidelines and an emphasis on transparency within the HCI community, evidence suggests that these efforts have not significantly influenced the methodological detail in conventional journal publications~\cite{niksirat2023changes, echtler2018open, clayson2019methodological, paul2022preprocessing}. The lack of methodological detail indicates a potential need for alternative strategies to enhance transparency and reporting in the HCI field.

As a result, these recent surveys imply that understanding and replicability issues persist in HCI. Therefore, it is time to establish more intensive cross-disciplinary conversations to synthesize current practices in physiological HCI research. This workshop will combine quantitative and qualitative researchers working with biodata to address issues around the open science, use and interpretation, and ethics of physiological data in HCI. Ultimately, we will deliver a living document formalizing best practices as in previous work \cite{babaei2021critique} (see, \url{https://edaguidelines.github.io/}) to provide the HCI community with sound and robust guidelines.


\subsection{Workshop Topics}
\label{wk_topics}
We aim to structure group discussions during this workshop around \textbf{four key topics and challenges} commonly faced by researchers and designers working with physiological signals. The discussion will allow attendees to share their experiences and help to identify and articulate common challenges across diverse domains. Through this communal effort, we will begin to map out future directions for the field of physiological computing to address these challenges.

\subsubsection{Topic 1: Replicability and Research Transparency.}
As HCI continues integrating physiological signals, applying FAIR Principles\footnote{\url{https://www.go-fair.org/fair-principles/}} becomes necessary for repeatability, reproducibility, and replicability. Several processes and platforms, like the Open Science Framework\footnote{\url{https://osf.io/}} and Zenodo\footnote{\url{https://zenodo.org/}}, have been developed to foster community building. Open data-sharing enables result validation, supports data aggregation for meta-analysis, encourages creative re-analysis, and adapts to evolving scientific methodologies. 
However, open data alone does not guarantee replicability. Often, data from individual, small-scale research projects --termed the "long tail of science"-- is produced with specific contexts or constraints in mind~\cite{ferguson2014big}. To make this data usable by the wider community, there is a need for standardized documentation or even formal models detailing how the data should be shared and interpreted. 
For instance, OpenNeuro\footnote{\url{https://openneuro.org/}} or ARTEM-IS~\cite{vsovskic2023artem} advocate for standardized datasets with joint documentation models. Such efforts have demonstrated that data sharing can amplify scientific studies' impact, drawing contributions from various disciplines~\cite{vicente2018open}. In the HCI domain, there are notable examples like \citet{babaei2021critique} practices for EDA recording and \citet{bergstrom2021evaluate} checklist for VR experiments. The Association for Computing Machinery (ACM) introduced badges\footnote{\url{https://www.acm.org/publications/policies/artifact-review-badging}} to reward research transparency, acknowledging the broader movement toward open science.

While there's been a push towards reproducibility and transparency, e.g., the first HCI conference introduced an open science track\footnote{\url{https://iui.acm.org/2023/call_for_open_science.html}}, or with changes even in the review processes of prominent HCI conferences, studies suggest that many in the field still do not openly publish their research artifacts~\cite{niksirat2023changes}. This could stem from misunderstandings about the process, the multifaceted nature of HCI, or underestimating the importance of transparency. Balancing privacy concerns with open science ambitions remains a complex challenge, especially when dealing with sensitive physiological data.

\subsubsection{Topic 2: I/O of Physiological Signals.}
Integrating physiological signals in HCI presents many technical challenges~\cite{Schmidt2016}. At the core, issues are related to the robustness, scalability, and adaptability of continuously sensing (wearable) physiological sensors~\cite{Tucker2022}. Especially outside the controlled environment of a lab, these physiological sensors often demand calibration and are vulnerable to motion artifacts~\cite{lopes2021physiological}. This is further complicated when systems rely on physiological data to infer users' states or emotions, such as anxiety or task engagement, as discussed in our previous workshops \cite{el2020meec, el2021meec}. Concerning scalability, researchers are increasingly adopting software-based physiological measurements, such as pulse and vital sign measurement using remote PPG from facial data, and used across mixed reality setups~\cite{McDuff2017}. To advance the field, it is essential to go beyond isolated use cases and establish foundational physiological principles. Further challenges include coupling physiological signals with the appropriate body feedback modalities~\cite{Alfaras2020, chiossi2022adapting} or body augmentation, which can lead to questioning the sense of agency~\cite{Cornelio2022}.

\subsubsection{Topic 3: Meaning-Making and User Experience of Physiological Signals}
The various applications using biosignals reveal the plurality of effects this technology can have on individuals and our relationship to our own and others' bodies. These effects can have long-term consequences of reshaping these relationships beyond the moment of the interaction. For instance, using biofeedback can improve one's interoceptive awareness (awareness of one's internal processes)~\cite{lehrer2013dynamic}, and receiving information about others' internal state can improve users' ability to empathize with them, ultimately enhancing mutual understanding~\cite{dey2019sharing, hirsch2023heart}. This could be particularly useful when only limited affective cues are shared between users, e.g., when people lack sensitivity to such cues, such as users with autism. However, this enhanced access to others' states through biosignals also carries the risk of drawing too much attention towards an externalized representation of biodata and away from the human at the other end.

Biodata provides us with insights into the state and process of our bodies. However, interpreting these signals is often ambiguous~\cite{howell2016biosignals}. For example, what does it mean if my heart rate is 10 beats faster than my partner's, and how does this influence biofeedback displays (cf.,~\cite{el2023my})? Considering a case from affective computing, there is a growing debate arising from the constructionist models of emotions~\cite{lindquist2022cultural} on how much could be possible to infer from the physiological data about users' emotions given the lack of universal expressions of emotions~\cite{barrettAIWeighsDebate2021}. The influence of the context of use presents another challenge for considering variant meaning-making processes. Naturally, our social relationships significantly affect our comfort level with sharing our intimate biodata. A heartbeat visualization that may be intimate for close relationships~\cite{liu2021significant}, it might feel like oversharing in a professional relationship~\cite{qin_having_2020}, or can represent an abstract human unity when received from a distant stranger~\cite{lozano-hemmer2019pulse}. 
\vspace{-5pt}

\subsubsection{Topic 4: Ethical and Privacy Concerns} 
Working with biosignals inevitably raises ethical concerns and privacy considerations. Our physiology is inherently private, and by that nature, so should biodata be considered sensitive and private, providing agency to each individual to make an informed decision to share. In affective computing, an ethics framework has been proposed where recommendations are given to developers, operators, and regulators~\cite{ongEthicalFrameworkGuiding2021}. This is because of the unexpected usages that arise beyond the expectations of the initial system designer.

Biosignals often disclose insights into our internal states beyond our immediate awareness, complicating the issue of consent for data sharing. Our limited grasp over such physiological processes underscores the need for establishing transparent, meaningful ground truths to prevent misinterpretations and ensure ethical use in biosignal research. Hence, its interpretive potential when making it accessible to others. Thus, we must consider users' sensitivity and potential vulnerability when designing biodata-sharing systems. Questions of privacy, agency over data, equity and power relationships, data use, and storage are crucial in considering how we design such technology. This issue is further amplified by the lack of consistent data privacy and security standards across health, academia, and private business domains, where the industry is often not bounded by the same data security standards as academic and medical fields. Could a privacy-by-design approach be an initial solution? With this approach, we can embed privacy considerations into every stage of adaptive and biofeedback systems' design and development process~\cite{chiossi2023can}. Alternatively, federated learning could offer a solution by enabling model training without centralizing sensitive physiological data, ensuring privacy and regulatory compliance~\cite{ju2020federated}.

\section{Organizers}
\textbf{Francesco Chiossi} (\url{https://www.francesco-chiossi-hci.com/}) is a PhD researcher in the Media Informatics Group at LMU Munich. With a background in cognitive neuroscience, he focuses on implicit measures of human behavior, such as EDA and EEG, as an implicit input to design physiologically-adaptive systems.

\textbf{Ekaterina R. Stepanova} is a Ph.D. Candidate at the School of Interactive Arts and Technology at Simon Fraser University with a background in cognitive science, developmental psychology, and virtual reality. Her research employs somaesthetics and embodied cognition to design mediated experiences with bioresponsive and immersive technologies.

\textbf{Benjamin Tag} is a Lecturer at Monash University. He researches Human-AI Interaction, Digital Emotion Regulation, and human cognition with a focus on inferring mental state changes from physiological data collected in the wild.

\textbf{Monica Perusquia-Hernandez} (\url{https://www.monicaperusquia.com/}) is an assistant professor at the Nara Institute of Science and Technology (NAIST), Japan, working in affective computing, signal processing, and interoceptive awareness enhancement in cyber-physical systems. Her work relies on Computer Vision, EMG, EEG, ECG, and EDA for congruence estimation between facial expressions and emotions.   

\textbf{Alexandra Kitson} (\url{https://www.alexandrakitson.com}) is a postdoctoral fellow in the Tangible Embodied Child Computer Interaction Lab at Simon Fraser University whose research is focused on designing, developing, and evaluating interactive systems such as wearables and virtual reality to support both personal and social transformation and emotional well-being.

\textbf{Arindam Dey} is a computer scientist on a mission to make Metaverse better for users in various ways. Currently, he is a Research Scientist at Meta, focusing on health and safety in the metaverse. He is also Honorary Research Fellow at the University of Queensland, Australia, primarily focusing on Mixed Reality and Empathic Computing.

\textbf{Sven Mayer} (\url{https://sven-mayer.com}) is an assistant professor at LMU Munich. His research sits at the intersection between HCI and Artificial Intelligence, where he focuses on the next generation of computing systems. He uses artificial intelligence to design, build, and evaluate future human-centered interfaces.

\textbf{Abdallah El Ali} (\url{https://abdoelali.com}) is an HCI research scientist at Centrum Wiskunde \& Informatica (CWI) in Amsterdam within the Distributed \& Interactive Systems group. He leads the research area on Affective Interactive Systems, combining advances in HCI, eXtended Reality, and Artificial Intelligence to measure, infer, and augment human cognitive, affective, and social interactions.

\section{Plan To Publish Proceedings}
We will publish the workshop proceedings on \href{http://ceur-ws.org/}{CEUR-WS.org}. Moreover, the workshop website will host papers accepted two weeks before the conference upon the authors' consent.

\section{Website}
The website will contain a call for papers, links to materials and activities for asynchronous engagement, and links to accepted position papers.  Finally, the most important contribution of the website will be a living document that relates to best practices for biosignals research in HCI concerning validity, reproducibility, and privacy.

\section{Pre-Workshop Plans:} 

The plan for this workshop began at DIS~\cite{stepanova2023designing} and at MobileHCI~\cite{schneegass2023future}. From those two workshops, we have established a community and joint effort on Slack\footnote{\href{https://join.slack.com/t/physiological-kho7240/shared_invite/zt-24gg6nras-NsaVl39oMMLBpwBooZ7E2Q}{Physiological Interaction Slack Server}}. 
Organizing this workshop is the important next step in the long-term plan discussed, and members of the new micro-community on Slack were invited to contribute to the organization of this new workshop. A dedicated webpage will be hosted on the first author website\footnote{\url{https://www.hcilab.org/physiochi24}}. We will promote the workshop on the Slack server, in research groups, and at upcoming HCI conferences. Standard CfP releases via mailing lists and social media channels will also be used to increase the reach and inclusivity of the event.

To ensure a productive discussion, we will select participants with relevant expertise working with biosignals from a physiological computing, design, and ethics perspective. Prospective participants will be invited to either submit a position paper (1-4 pages), or a copy of their previously published paper exploring one of the workshop topics. Authors will be encouraged to follow the accessibility guidelines for their submissions. Additionally, participants will be asked to complete a short survey to help us better understand the distribution of participants and forms of engagement (online or in-person) to optimize the planned workshop structure. The survey will inquire on the planned form of attendance, the primary field or work, ranking of the interest in proposed workshop topics with an option to suggest new ones, and consent to be invited to a Slack group for communication with organizers and other attendees before the workshop. We aim to attract and select about 30-35 participants. 

\paragraph{Review of Submissions}
Our focus for reviewing will be on the submissions' potential to provoke relevant discussion at the workshop. We expect position papers that present provocative views of the future, case studies, or extensions to existing work to discuss interesting research outcomes. The workshop organizers will primarily be responsible for reviewing and determining acceptance. If we receive unrelated submissions, this will be expanded to those in the existing Slack community.

\section{Workshop Mode}
We will conduct a synchronous hybrid workshop. Our hybrid approach will facilitate sharing between virtual and in-person attendees, ensuring discussion grounded in equitable and diverse participation, integrating a broad range of perspectives. 

\paragraph{Hybridity and Asynchronous Engagement}
During the workshop, we will introduce discussion topics to attendees using a presentation projected in the physical room and shared on the teleconference platform for remote attendees. We will use a Miro board to note discussion topics to make them available for in-person and remote participants. One of the organizers will copy the topics from the whiteboard into the Miro board and vice-versa. Remote participants will use the Miro Boards, which will also be projected on the whiteboards in the conference room. This way, participants in the conference room can add physical notes to the same discussion space and see the contributions of remote participants. We will use automated transcription in Zoom during large group discussions to improve accessibility. We will share all the materials on the workshop website and through the Slack channel for access by participants who cannot attend synchronous sessions. These materials will include papers submitted by accepted participants, descriptions of the provocation exercises that participants can experiment with by themselves, prompt cards used for structuring the discussion, and links to the Miro board summarizing our discussion.  

\paragraph{Materials} We will bring sketching materials, such as paper, sticky notes, markers, etc., for physically present participants to note down their discussion and prepare a Miro board for online participants and the summary of everyone's discussion. 
Website, Miro board, and all other materials will adhere to accessibility criteria outlined by ACM. We will continue to consult with CHI Accessibility Chairs to ensure accessibility before and during the workshop.

\paragraph{Accessibility and Inclusivity}
Accepted authors will be expected to enhance the accessibility of their submissions by including accurate subtitles for videos and ensuring their PDFs are screen-reader-friendly. We will proactively contact our workshop attendees to identify and address additional accessibility requirements for the event day.

\section{Workshop Structure}
The workshop spans a \textbf{full day} and will be built into four rounds (arranged around the natural breaks in the conference). In all parts, the aim is to encourage discussion, especially before breaks and lunch, where the most natural discussions are likely to occur.  One organizer will lead the in-person workshop for each program section, ensuring a lively and interactive session. Simultaneously, a second organizer will facilitate the online participation and discussion, ensuring that remote participants are equally engaged and included in the proceedings.

\paragraph{\textbf{Round 1: Kickoff and Speed Dating (40 minutes)}} The workshop will begin with an introduction by the organizers and a short warm-up speed-dating activity. This speed-dating activity will help attendees to focus on the workshop interest and engage with the other participants. Participants will be asked to introduce themselves and explain their motivation to participate in the workshop. They will also be asked to briefly introduce the work (e.g., method or application). Due to the expected diversity in participants' research backgrounds, we will start by organizing into different groups based on identifying statements (e.g., qualitative researcher or quantitative researcher) based on the survey we ask participants to fill in. The ultimate aims of Round 1 are a) to explicate the scope of the workshop and the expertise in the room, b) to highlight the variety of expertise, and c) to end up with mixed groups around the tables. Further, by doing so, we aim to avoid being on laptops and settling into a passive form of listening to talks. Once in mixed groups, the remainder of R1 will focus on important shared research questions and create a physical post-it mindmap on an available large surface and a Miro board for online attendants. 

\paragraph{\textbf{Round 2: Research Transparency for Biosignals in HCI (60 minutes)}} Once in mixed groups, in R2, we will focus on important challenges for Open Science practices and create a physical post-it mindmap or on a Miro Board. Chiossi will provide an introductory talk (10 minutes) on open guiding principles~\cite{wilkinson2016fair}, where data should be Findable, Accessible, Interoperable, and Reusable (FAIR). This allows participants to have a shared background in the topic. Then, participants will engage in 3 rounds of medium to small group discussions (4-8 participants per group) in the form of roundtables. The main goal of each group is to \textit{identify how each group member follows or does not follow FAIR guidelines in their work}. One moderator per group will be chosen to report to all other participants on the discussion outcomes. Each group, online and offline, will be joined by one organizer, who will focus on timekeeping and deliverables to ensure that the groups are ready to contribute to the large group discussion at the end. We expect to discuss all FAIR principle statements before lunch. This exercise will flow naturally into the first coffee break, where discussions can be continued. 

\paragraph{\textbf{Round 3: Keynote. (60 minutes)}} We plan to invite one keynote speaker to the workshop. The keynote will be an established researcher in the physiological computing area, and the talk will cover all four topics of the workshop in 30 minutes, followed by an extensive Q \& A and a discussion session. This round will have an expected duration of one hour.

\paragraph{Discussion Lunch (90 minutes)} Depending on the arrangements available for lunch in the area, we aim to send groups (arranged in R1) to lunch to continue discussions. We will ask each group to return to the afternoon session with a considered set of biosignal measures that they employ in their research and which challenges they face regarding the construct validity. The lunch session is expected to last 90 minutes.

\paragraph{\textbf{Round 4: Meaning-Making and UX of Biosignals (45 minutes)}}  To avoid a post-lunch slump, this session will first bring the ideas back to the room from lunch and start with grouping researchers by biosignal of interest: brain (EEG and fNIRS), peripheral (EDA, ECG, electrogastrography), biosignals with implicit and explicit control (eye-tracking and electromyography, i.e., EMG). Each group will develop a shared definition of how each biosignal is physiologically interpreted, and to ensure a breadth of discussion, list what each signal is mapped in their HCI research to which construct and as input for interaction. The definition and application areas will be presented and discussed by a representative of each group. 

\paragraph{\textbf{Round 5: Sharing Biosignals (45 minutes)}}

Given the now shared theoretical background and alignment on definitions across participants, we will switch the focus from physiological evaluation to a multi-user, social perspective of sharing biosignals. Here, participants who selected the topic of sharing biosignals will pitch their presentations on their submitted position paper - case study. Groups previously formed in Round 4 will discuss how biodata can be represented, either individually~\cite{hirsch2023heart} or as an aggregate from multiple users, offering potential privacy solutions~\cite{qin_having_2020}. Second, groups will discuss the centrality of visual representations in biofeedback designs~\cite{lux2018live} and explore the potential of other sensory modalities. Participants will highlight the challenges and opportunities in mapping and representing signals for multiple users, presenting the outcome of discussions by emphasizing the need for symmetrical data representation and its impact on user interpretation. 

\paragraph{\textbf{Round 6: Privacy and Ethics (60 minutes)}} The final part of the afternoon block activity will focus on responsible research, ethics, and privacy. As a community, we expect this to be a key focus issue for the future, with physiological interaction and technology presenting several potentially invasive challenges. We intend to use the Legal and Moral-IT cards on tables to provoke discussion\footnote{\url{https://lachlansresearch.com/the-moral-it-legal-it-decks/}}. Either submitting authors or invited speakers will be invited to co-facilitate this activity, where they have already considered aspects of this theme. 

\section{Post-Workshop Plans}

We will publish the guidelines summary on the workshop website. Ideally, this could result in a joint publication between organizers and participants.  Finally, we will prepare a proposal for a special issue at TOCHI on integrating biosignals in HCI, potentially including a Dagstuhl proposal and further workshops at SIGCHI conferences (IUI and UbiComp). Additionally, involved authors will mentor early-career researchers via the Slack Server and future events. The Slack Server will serve as a platform to engage participants, continuing the evolving discussion of challenges and opportunities presented by designing with biosignals: post questions, share posts about their work, recruit participants, and form collaborations. 

\section{Call for Participation}
Biosensing technologies are increasingly more widely integrated in HCI. Biosignals provide novel opportunities for interaction, offering valuable insights into ordinarily hidden processes inside our bodies, revealing somatic information about our and other's bodies, emotions, health, and cognitive processes. However, integrating biosignals in HCI presents many challenges about transparency, UX, I/O, interpretation of biodata, and broader ethical concerns. To map out the landscape of existing challenges and future research directions, we invite participants working with biosignals to join a one-day hybrid workshop held at the 2024 ACM SIGCHI Conference on Human Factors in Computing Systems following a hybrid format. We welcome participants from HCI, Psychology, Neuroscience, Data Science, Mixed Reality, and Digital Ethics.
We invite submissions of 2-4 page position papers or case studies, presenting a project, or articulating a challenge related to this call. Alternatively, authors may submit their previously published papers raising relevant questions. Submissions should be sent to EasyChair \footnote{\url{https://easychair.org/conferences/?conf=physiochi24}}, must adhere to accessibility guidelines outlined by ACM, and use a CEUR Workshop Proceedings template. The organizing committee will select submissions based on the quality and contribution of the work relating to the workshop themes, with a special focus on biosignal integration. At least one author of each accepted submission must attend (in-person or remote) and register for the workshop and the CHI’24 conference. For more information, please visit \url{https://www.hcilab.org/physiochi24/}.

\begin{acks}
Francesco Chiossi was supported by the Deutsche Forschungsgemeinschaft (DFG, German Research Foundation), Project-ID
251654672-TRR 161. Katerina Stepanova was supported by the Social Sciences and Humanities Research Council of Canada. M. Perusquia-Hernandez was supported by JSPS Kakenhi Grant Number 22K21309. Sven Mayer was supported by National Research Data Infrastructure for and with Computer Science (NFDIxCS).
\end{acks}

\bibliographystyle{ACM-Reference-Format}
\bibliography{bibliography}


\end{document}